\documentclass[aps,prb,preprint,superscriptaddress]{revtex4-1}

\usepackage[]{typearea}

\usepackage[utf8]{inputenc}
\usepackage{graphicx}
\usepackage[ngerman,english]{babel}
\usepackage{natbib}
\usepackage{color}
\usepackage{siunitx}
\usepackage{calc}

\usepackage{amssymb}
\usepackage{amsmath}
\usepackage{amsfonts}

\usepackage{textcomp}
\usepackage{booktabs}
\usepackage{subcaption}

\begin{document}


\title{Nanowire magnetic force sensors fabricated by focused electron
	beam induced deposition}



\author{H.~Mattiat} \affiliation{Department of Physics, University of
	Basel, 4056 Basel, Switzerland} \affiliation{Swiss Nanoscience
	Institute, University of Basel, 4056 Basel, Switzerland}

\author{N.~Rossi} \affiliation{Department of Physics, University of
	Basel, 4056 Basel, Switzerland} \affiliation{Swiss Nanoscience
	Institute, University of Basel, 4056 Basel, Switzerland}

\author{B.~Gross} \affiliation{Department of Physics, University of
	Basel, 4056 Basel, Switzerland} \affiliation{Swiss Nanoscience
	Institute, University of Basel, 4056 Basel, Switzerland}

\author{J.~Pablo-Navarro} \affiliation{Laboratorio de Microscop\'ias
	Avanzadas (LMA), Instituto de Nanociencia de Arag\'on (INA),
	Universidad de Zaragoza, 50018 Zaragoza, Spain}
\affiliation{Instituto de Ciencia de Materiales de Arag\'on and
	Departamento de F\'isica de la Materia Condensada, CSIC-Universidad
	de Zaragoza, 50009 Zaragoza, Spain}

\author{C.~Mag\'{e}n} \affiliation{Laboratorio de Microscop\'ias
	Avanzadas (LMA), Instituto de Nanociencia de Arag\'on (INA),
	Universidad de Zaragoza, 50018 Zaragoza, Spain}
\affiliation{Instituto de Ciencia de Materiales de Arag\'on and
	Departamento de F\'isica de la Materia Condensada, CSIC-Universidad
	de Zaragoza, 50009 Zaragoza, Spain}

\author{R.~Badea} \affiliation{Department of Physics, Case Western
	Reserve University, Cleveland, Ohio 44106, USA}

\author{J.~Berezovsky} \affiliation{Department of Physics, Case
	Western Reserve University, Cleveland, Ohio 44106, USA}

\author{J.~M.~De Teresa} \affiliation{Laboratorio de Microscop\'ias
	Avanzadas (LMA), Instituto de Nanociencia de Arag\'on (INA),
	Universidad de Zaragoza, 50018 Zaragoza, Spain}
\affiliation{Instituto de Ciencia de Materiales de Arag\'on and
	Departamento de F\'isica de la Materia Condensada, CSIC-Universidad
	de Zaragoza, 50009 Zaragoza, Spain}

\author{M.~Poggio} \affiliation{Department of Physics, University of
	Basel, 4056 Basel, Switzerland} \affiliation{Swiss Nanoscience
	Institute, University of Basel, 4056 Basel, Switzerland} \email{martino.poggio@unibas.ch}


\date{\today}

\begin{abstract}
We demonstrate the use of individual magnetic nanowires (NWs), grown
by focused electron beam induced deposition (FEBID), as scanning
magnetic force sensors.  Measurements of their mechanical
susceptibility, thermal motion, and magnetic response show that the
NWs posses high-quality flexural mechanical modes and a strong
remanent magnetization pointing along their long axis.  Together,
these properties make the NWs excellent sensors of weak magnetic
field patterns, as confirmed by calibration measurements on a
micron-sized current-carrying wire and magnetic scanning probe
images of a permalloy disk. The flexibility of FEBID in terms of the
composition, geometry, and growth location of the resulting NWs,
makes it ideal for fabricating scanning probes specifically designed
for imaging subtle patterns of magnetization or current density.
\end{abstract}


\maketitle


\section{Introduction}
\label{sec:introduction}

In the early 1800s, images of the stray magnetic fields around
permanent magnets and current-carrying wires made with tiny iron
filings played a crucial role in the development of the theory of
electromagnetism.  Today, magnetic imaging techniques such as Lorentz
microscopy, electron holography, and a number of scanning probe
microscopies continue to provide invaluable insights.  Images of
magnetic skyrmion configurations~\cite{yu_real-space_2010} or of edge
and surface currents in topological
insulators~\cite{nowack_imaging_2013} have provided crucial direct
evidence for these phenomena.  The ability to map magnetic field
sensitively and on the nanometer-scale -- unlike global magnetization
or transport measurements -- overcomes ensemble or spatial
inhomogeneity in systems ranging from arrays of nanometer-scale
magnets, to superconducting thin films, to strongly correlated states
in van der Waals heterostructures.  Local imaging of nanometer-scale
magnetization~\cite{thiel_probing_2019}, local Meissner
currents~\cite{jelic_imaging_2017}, or current in
edge-states~\cite{uri_nanoscale_2019} is the key to unraveling the
microscopic mechanisms behind a wealth of new and poorly understood
condensed matter phenomena.

The techniques combining the highest magnetic field sensitivity with
the highest spatial resolution include scanning Hall-bar microscopy,
scanning nitrogen-vacancy (NV) center magnetometry, and scanning
superconducting quantum interference device (SQUID) microscopy.  Each
has demonstrated a spatial resolution better than
\SI{100}{\nano\meter} and a magnetic field sensitivity ranging from
$\SI{500}{\micro\tesla}/\sqrt{\SI{}{\hertz}}$ for Hall-bar
microscopy~\cite{hicks_noise_2007}, to
$\SI{60}{\nano\tesla}/\sqrt{\SI{}{\hertz}}$ for NV
magnetometry~\cite{maletinsky_robust_2012}, and
$\SI{5}{\nano\tesla}/\sqrt{\SI{}{\hertz}}$ for scanning SQUID
microscopy~\cite{vasyukov_scanning_2013}.  Recently, a form of
magnetic force microscopy (MFM) based on a transducer made from a
magnet-tipped nanowire (NW) demonstrated a high sensitivity to
magnetic field gradients of
$\SI{11}{\milli\tesla}/(\SI{}{\meter} \sqrt{\SI{}{\hertz}})$ with a
similar spatial resolution~\cite{rossi_magnetic_2019}.  The high force
sensitivity of NW cantilevers coupled together with a small magnetic tip size
could allow such sensors to work both close to a sample, maximizing
spatial resolution, and in a regime of weak interaction, remaining
noninvasive.

Here, we demonstrate the use of individual magnetic NWs, patterned by
focused electron beam induced deposition (FEBID), as MFM transducers
for mapping magnetic fields with high sensitivity and resolution.  The
monopole-like magnetic charge distribution of their tips makes these
transducers directly sensitive to magnetic fields rather than to field
gradient, as in the initial demonstration of NW
MFM~\cite{rossi_magnetic_2019}.  Furthermore, the FEBID fabrication
process allows for a large degree of flexibility in terms of the
geometry, composition, and location of the NW transducers.  In
particular, the possibility of long, thin, and sharp NWs is promising
for further increasing field sensitivity and spatial resolution of the
technique~\cite{braakman_force_2019}.

\section{FEBID NWs}
\label{sec:FEBID}

FEBID is an additive-lithography technique where precursor gas
molecules are adsorbed onto a surface and dissociated by a focused
electron beam, forming a local
deposit~\cite{randolph_focused_2006,van_dorp_critical_2008,utke_gas-assisted_2008,huth_focused_2012,utke_nanofabrication_2012}.
It can be used to pattern exceptionally small features, down to a few
nanometers.  This high resolution patterning is complemented by the
capability to produce three-dimensional structures, as well as to
pattern on unconventional non-planar surfaces, such as
high-aspect-ratio tips.  FEBID and its sister technique, focused ion
beam induced deposition (FIBID), have been used to produce deposits of
various materials with metallic~\cite{gannon_focused_2004},
magnetic~\cite{wu_focused_2014,teresa_review_2016},
superconducting~\cite{sadki_focused-ion-beam-induced_2004}, or
photonic~\cite{esposito_nanoscale_2015} functionalities.  They have
been used in industry and research for mask
repair~\cite{bret_industrial_2014}, circuit editing, lamella
fabrication~\cite{giannuzzi_review_1999}, tip
functionalization~\cite{nanda_helium_2015}, and for the fabrication of
nano-sensors~\cite{schwalb_tunable_2010}.  They have also been
employed in the production of free-standing NWs from both
superconducting~\cite{cordoba_vertical_2018} and -- as in this work --
magnetic
materials~\cite{teresa_review_2016,pablo-navarro_tuning_2017}.

We grow free-standing NWs by FEBID using $\text{Co}_2(\text{CO})_8$ as
a gas precursor at specific positions along the cleaved edge of a
Au-coated GaAs chip. Their lengths range from \SIrange[range-phrase =
\text{ to }]{9.1}{11.0}{\micro\meter} and their base diameters from
\SIrange[range-phrase =
\text{ to }]{105}{120}{\nano\meter} as inferred from
scanning electron microscopy (SEM) images. They consist of
nanocrystalline Co, with a composition reaching up to
80\%~\cite{pablo-navarro_tuning_2017}, and residues of C and O.  Their
proximity to the edge of the chip allows optical access from the side
for the detection of their flexural motion.  A SEM image of a Co NW
standing at the chip edge is shown in Figure \ref{fig:nw}.  Surface
roughness and geometric irregularities are part of the FEBID
fabrication process and are present across the 11 NWs studied in this
work.

\section{Measurement setup}
\label{sec:setup}

\begin{figure}
	\centering
	\subcaptionbox{\label{fig:nw}\hfill\phantom{.}}{\includegraphics[height=0.3\textheight]{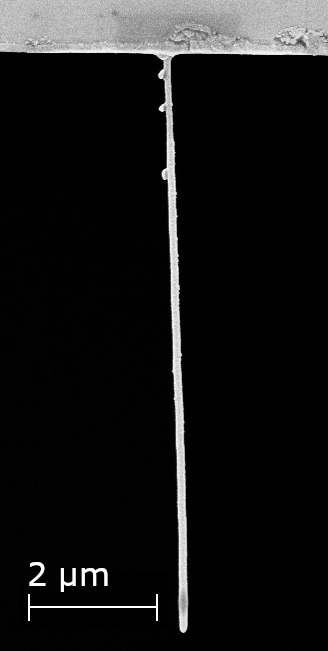}}%
	\hfill
	\subcaptionbox{\label{fig:modes}\hfill\phantom{.}}{\includegraphics[height=0.30\textheight]{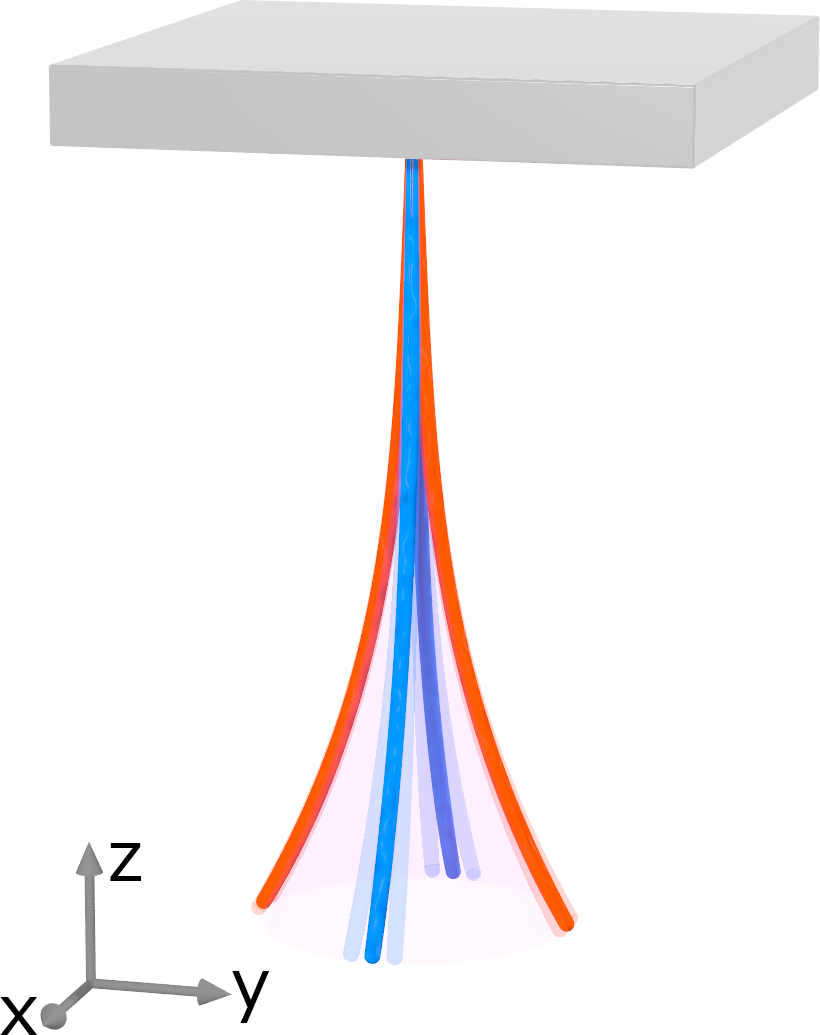}}
	\hfill
	\subcaptionbox{\label{fig:meas}\hfill\phantom{.}}{\includegraphics[width=0.28\textwidth]{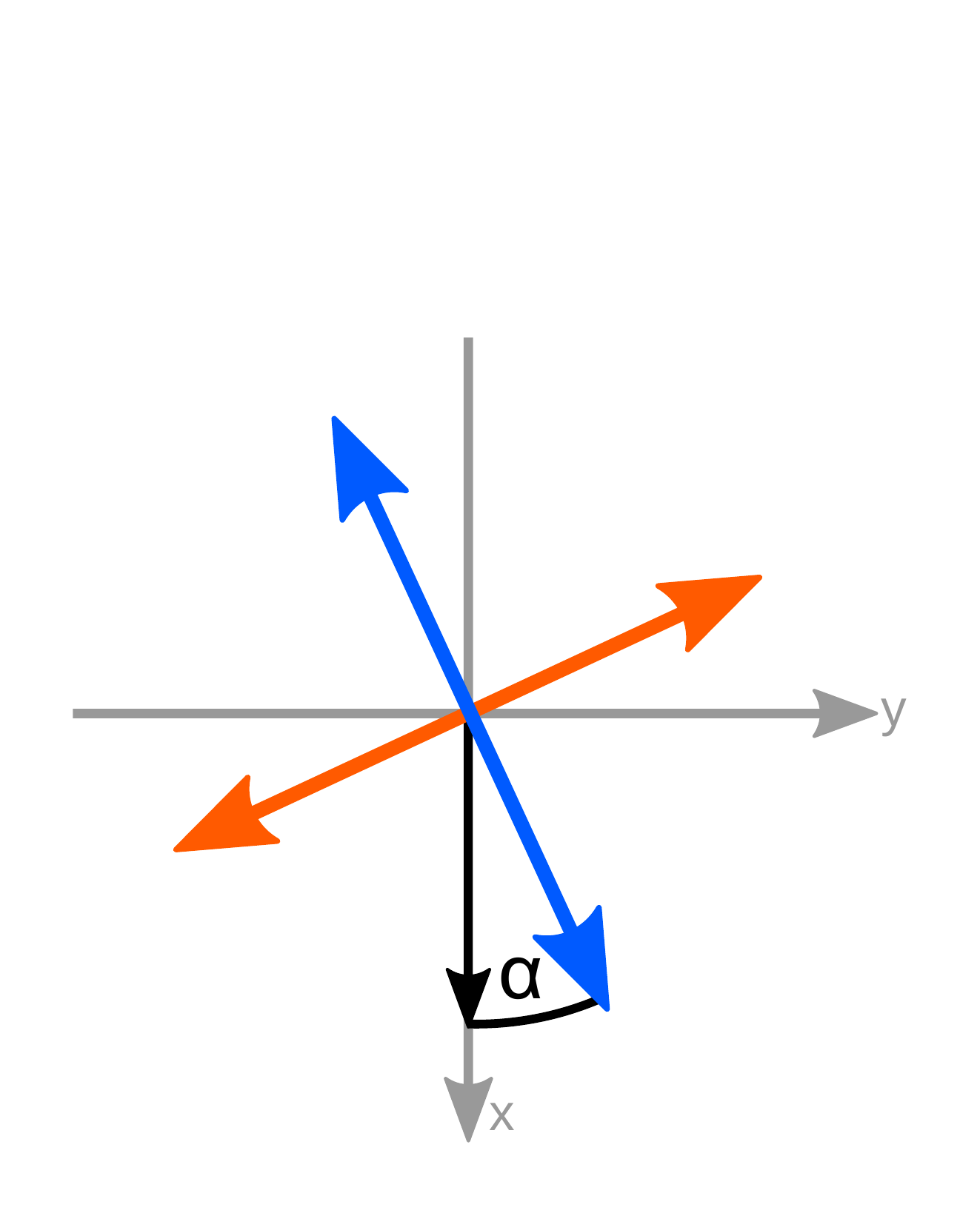}}%
	\\
	\vspace{0.4cm}
	\subcaptionbox{\label{fig:setup}\hfill\phantom{.}}{\includegraphics[width=0.47\textwidth]{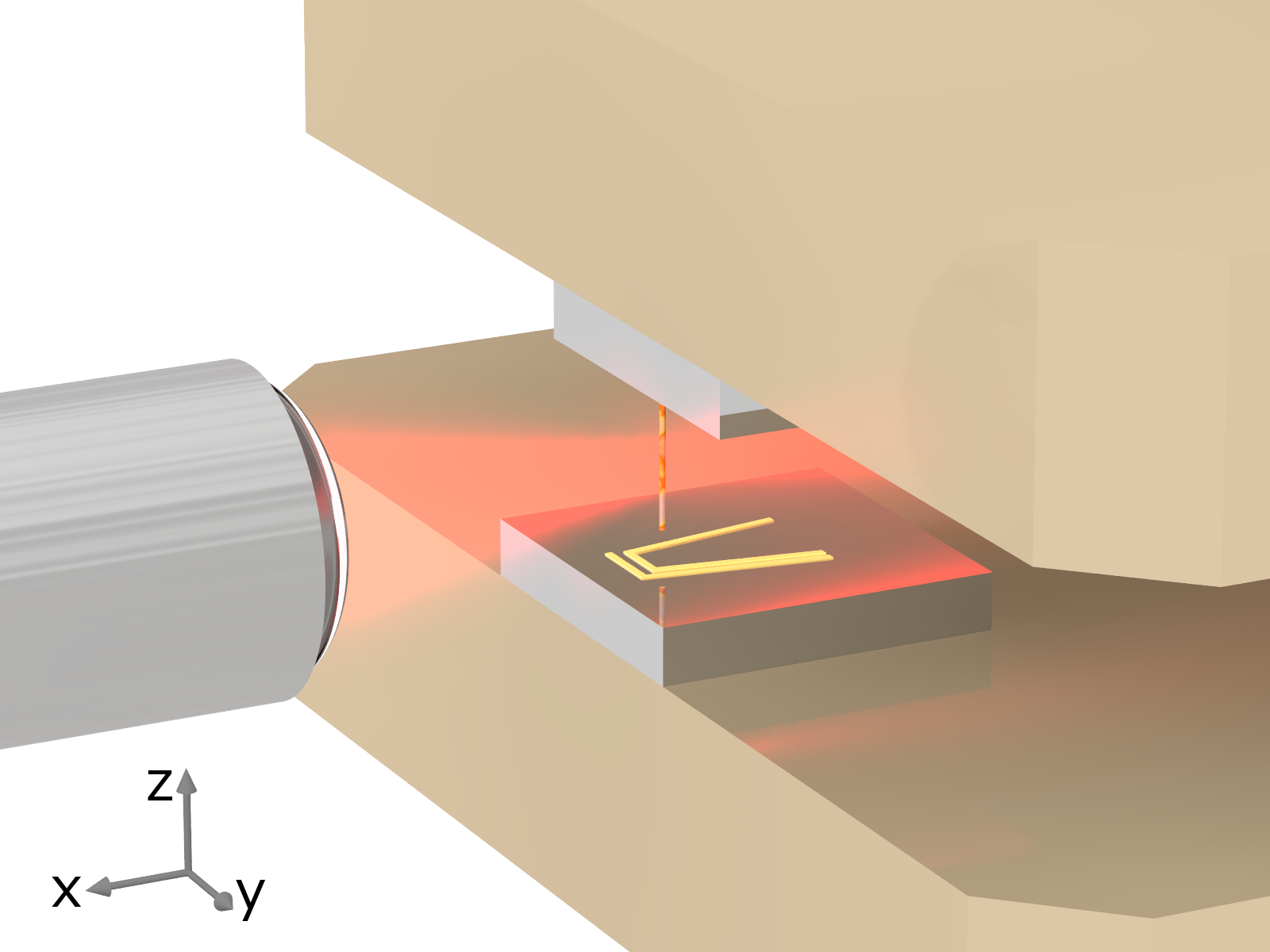}}%
	\hfill  
	\subcaptionbox{\label{fig:thnoise}\hfill\phantom{.}}{\includegraphics[width=0.47\textwidth]{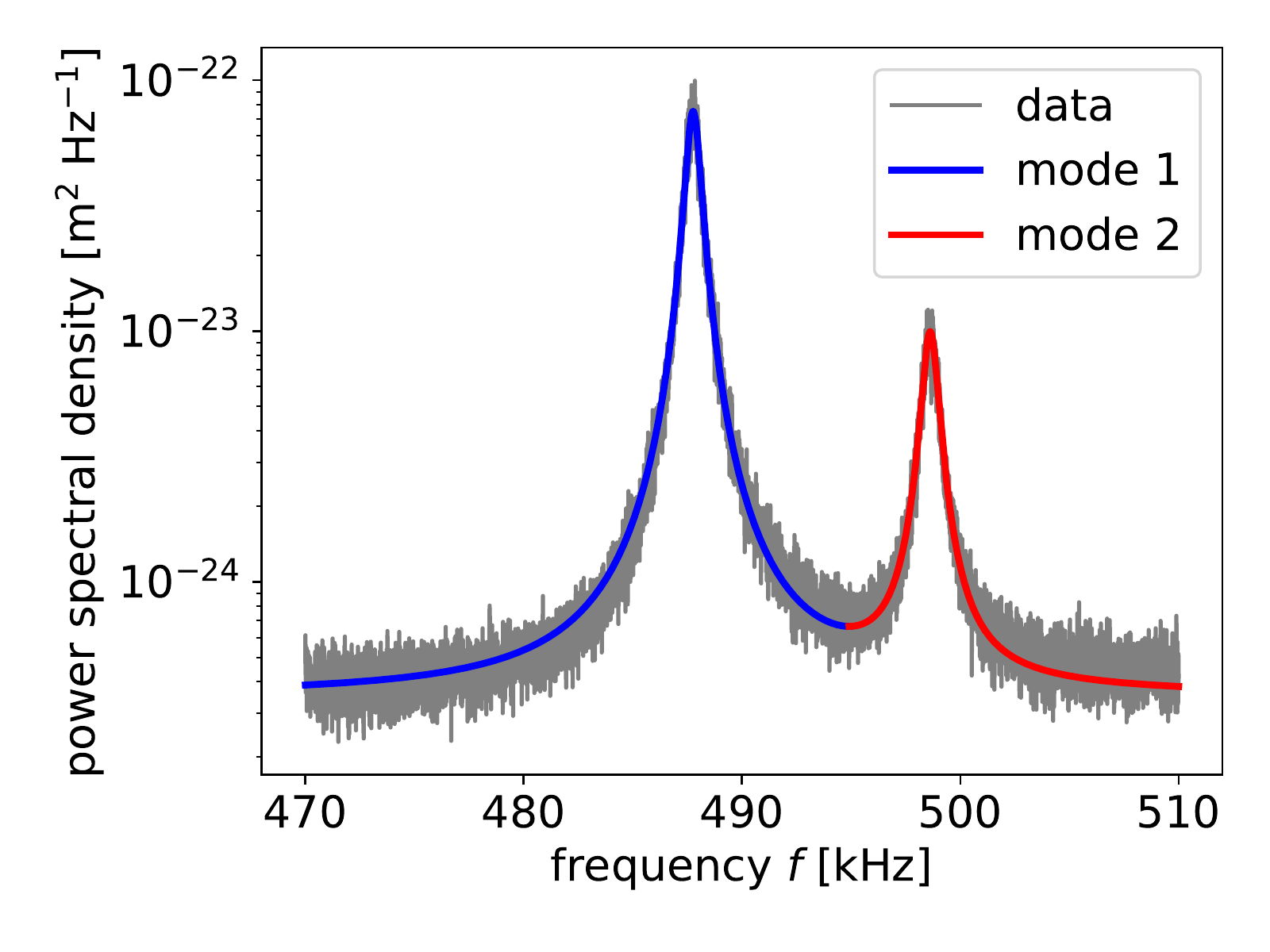}}%
	\caption{\textsf{\textbf{a}}) SEM micrograph of a Co FEBID NW
		at the chip edge. \textsf{\textbf{b}}) Illustration of the
		NW's flexural modes. The displacement amplitudes are
		exaggerated for better visibility. \textsf{\textbf{c}})
		Projection of flexural modes onto the $xy$-plane. $\alpha$
		is the angle between the lower frequency mode direction
		(blue) and the measurement vector
		(black). \textsf{\textbf{d}}) Illustration of the optical
		readout of the NW motion. \textsf{\textbf{e}}) Thermally
		excited response of the upper and lower mode at room
		temperature (gray). We fit the thermal noise displacement
		PSD with equation \eqref{eq:psd}. The two modes are
		highlighted by the blue and red regions of the
		fit. \label{fig:methods}}
\end{figure}

We grow free-standing NWs by FEBID using $\text{Co}_2(\text{CO})_8$ as
a gas precursor at specific positions along the cleaved edge of a
Au-coated GaAs chip. Their lengths range from \SIrange[range-phrase =
\text{ to }]{9.1}{11.0}{\micro\meter} and their base diameters from
\SIrange[range-phrase =
\text{ to }]{105}{120}{\nano\meter} as inferred from
scanning electron microscopy (SEM) images. They consist of
nanocrystalline Co, with a composition reaching up to
80\%~\cite{pablo-navarro_tuning_2017}, and residues of C and O.  Their
proximity to the edge of the chip allows optical access from the side
for the detection of their flexural motion.  A SEM image of a Co NW
standing at the chip edge is shown in Figure \ref{fig:nw}.  Surface
roughness and geometric irregularities are part of the FEBID
fabrication process and are present across the 11 NWs studied in this
work.

\captionsetup[subfigure]{margin=0pt, skip=0pt, textfont={sf}, position=top, labelformat=simple, labelfont={sf, bf}}

We mount the chip with as-grown Co NWs in a custom-built scanning
probe microscope, enclosed in a high-vacuum chamber at a pressure of
\SI{1e-6}{\milli\bar}.  The microscope includes a piezoelectric
translation stage, with which we position the NW of interest into the
focal spot of fiber-coupled optical interferometer for the detection
of the NW's flexural motion~\cite{nichol_displacement_2008}.  We use a
second piezoelectric translation stage to approach and scan the sample
of interest below the NW's free end, as illustrated in
Figure~\ref{fig:setup}.  This combined apparatus allows us to use
individual NWs as scanning probes operating in the pendulum geometry,
i.e.\ with their long axes perpendicular to the sample surface to
prevent snapping into
contact~\cite{rossi_vectorial_2017,rossi_magnetic_2019}.
\begin{table}
  \caption{Mechanical properties of representative NWs with the
    thermal bath held at room temperature
    ($T_{\text{bath}} = \SI{293}{\kelvin}$) and liquid helium
    temperature ($T_{\text{bath}} = \SI{4.2}{\kelvin}$) measured at
    high and low laser powers. Resonance frequencies, $Q$-factors,
    effective mass, and measurement angle (not shown here) are
    extracted by fitting the thermal noise PSD (see Appendix
    \ref{app:interferometer}). For a correct interpretation,
    bolometric heating due to the readout laser must be taken into
    account. Considering this effect, we also specify a NW temperature
    $T_\mathrm{NW}$ (see Supplementary Information). The data in the
    last two lines at liquid helium environment correspond to
    effective temperatures of \SI{99}{\kelvin} (\SI{20}{\micro\watt})
    and \SI{22}{\kelvin} (\SI{1.2}{\micro\watt})
    respectively. \label{tab:properties}}
\begin{tabular}{l*{7}{c}}
\bottomrule
\toprule
	$T_\mathrm{bath}$ &
	NW &
	$f_1$ [kHz] &
	$f_2$ [kHz] &
	$\Delta f_\mathrm{m}$ [kHz] &
	$Q_1$ &\textsl{}
	$Q_2$ &
	$m_{\mathrm{eff}}$ [\SI{e-15}{}kg]\\
\midrule
	293 & 1 & 390.726 & 426.018 & 35.292 & 528.0(1.4) & -- & 0.69(4)\\
	&4 & 514.459 & 556.750 & 42.291 & 551.1(1.5) & 527(66) & 0.260(15)\\
\midrule
	4.2 & 4 & 550.803 & 593.745 & 42.942 & 1045(1.0) & 1156(40) & 0.260(15)\\
	& 4 & 555.860 & 599.044(49) & 43.18 & 2356(39) & -- & 0.260(15)\\
\bottomrule
\toprule
\end{tabular}
\end{table}



The fiber-coupled optical interferometer operates at
\SI{1550}{\nano\meter} and provides a calibrated measurement of the
NWs flexural motion projected along the measurement axis (see Appendix
\ref{app:interferometer}).  Figure~\ref{fig:thnoise} shows a typical
power spectral density (PSD) of an individual NW's thermally excited
flexural motion at room temperature, revealing a splitting in
resonance frequency of the fundamental mode.  This well-known
splitting is observed for all examined NWs and is a signature of two
nearly degenerate, orthogonal flexural eigenmodes, resulting from
cross-sectional asymmetries and/or non-isotropic
clamping~\cite{cadeddu_time-resolved_2016}.  The NW's coupling to the
thermal bath results in a Langevin force that drives each mode
equally.  The difference in the amplitude of the two thermal noise
peaks in Figure~\ref{fig:thnoise} is a consequence of the projection
of the NW's flexural motion onto a single measurement axis,
corresponding to the direction of the optical gradient, at an angle
$\alpha$ with respect to mode 1.

\section{Mechanical properties}
\label{sec:mechproperties}

Measurements of the NWs' thermo-mechanical noise PSD are performed
with the bath held at room temperature (\SI{293}{\kelvin}), liquid
nitrogen temperature (\SI{77}{\kelvin}), and liquid helium temperature
(\SI{4.2}{\kelvin}).  Heating caused by absorption of the incident
\SI{25}{\micro\watt} laser light can increase the NW's temperature
well above the bath temperature.  As a result, care must be taken
interpreting PSDs, as discussed in the Supplementary Information.
Using the fits to the measured PSDs based on the
fluctuation-dissipation theorem, we determine the mechanical
properties of the fundamental flexural modes: their resonance
frequencies $f_i = \omega_i/2 \pi$, quality factors $Q_i$
($i = 1, 2$), and effective motional mass $m_{\text{eff}}$ (see
Appendix \ref{app:interferometer})~\cite{rossi_magnetic_2019}.  At
\SI{293}{\kelvin}, the resonance frequencies of the NWs are between
\SI{390}{\kilo\hertz} and \SI{560}{\kilo\hertz} with a mode splitting
from \SIrange{10}{42}{\kilo\hertz}. We measure quality factors around
600 and motional masses in the 100s of \SI{}{\femto\gram} range.
These parameters correspond to flexural modes with effective spring
constants $k_i$ of a few \SI{}{\milli\newton/\meter}. At a bath
temperature $T_{\text{bath}}=\SI{4.2}{\kelvin}$, the quality factors
improve by roughly a factor of 3 to around \SIrange{1000}{2000}{} and
the resonance frequencies shift upwards by roughly
\SI{30}{\kilo\hertz}. From these parameters, shown in Table
\ref{tab:properties} for two different NWs, we deduce the spring
constants, mechanical dissipation, and thermally limited force
sensitivities. Notably, at $T_{\text{NW}} = \SI{4.2}{\kelvin}$, a
typical NW has flexural modes with thermally-limited force
sensitivities around $\SI{10}{\atto\newton}/\sqrt{\SI{}{\hertz}}$. In
practice the force sensitivity is limited to about
$\SI{25}{\atto\newton}/\sqrt{\SI{}{\hertz}}$, since even at very low
laser power (\SI{1}{\micro\watt} on NW 4, signal-to-noise ratio of
first mode $\text{SNR} \approx 8$) bolometric heating is present and
leads to a NW temperature $T_\mathrm{NW} \approx \SI{20}{\kelvin}$
(see Supplementary Information).

\section{Magnetic properties}
\label{sec:magproperties}

\begin{figure}
	\centering
	\includegraphics[width=0.7\textwidth]{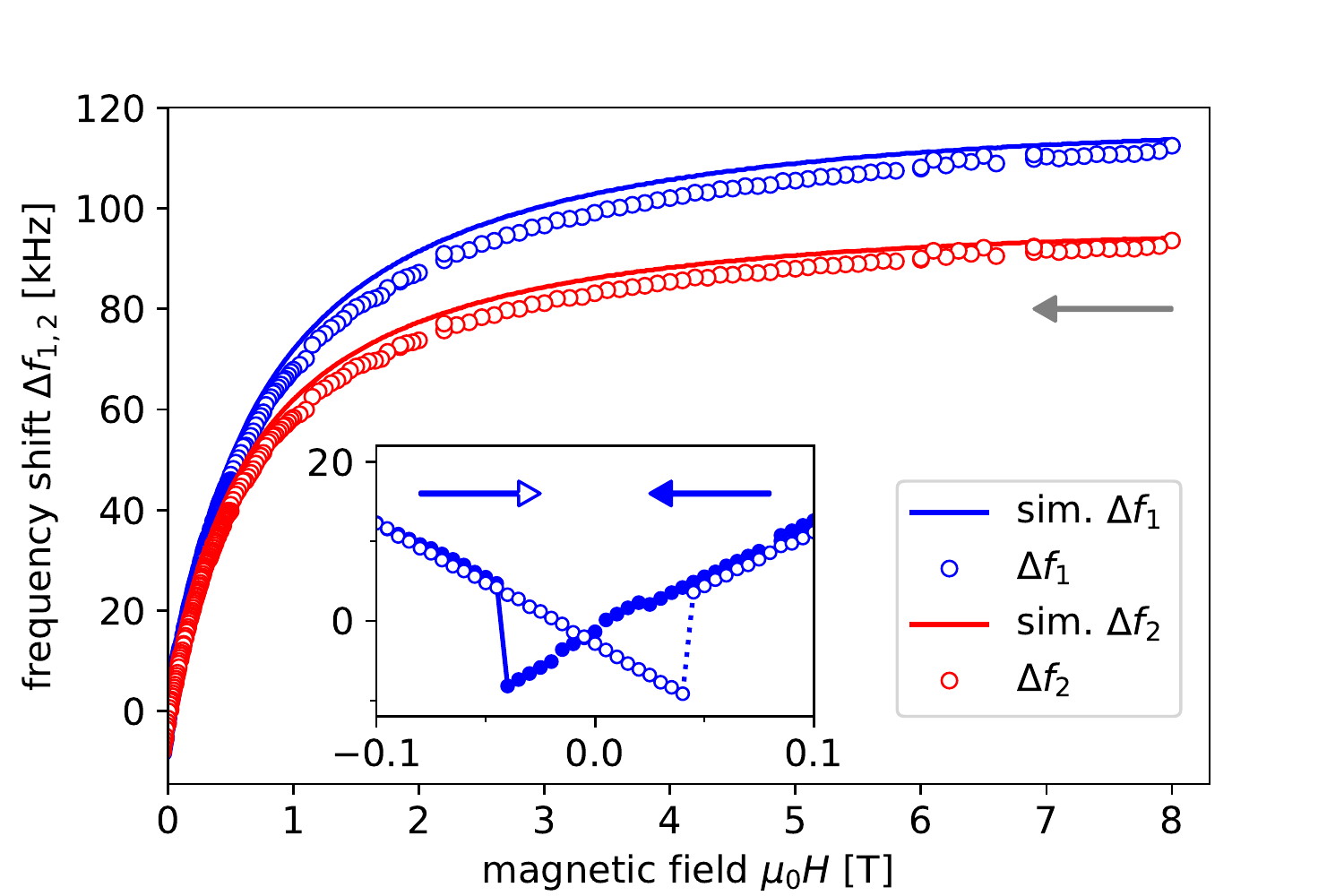}
	\caption{Resonance frequency shift of the two first order
          flexural modes of NW 4, measured while sweeping an axially
          aligned external magnetic field from \SIrange{+8}{0}{\tesla}
          (gray arrow). Points represent measured data, while solid
          lines correspond to simulations using \textit{Mumax3} (see
          Appendix \ref{app:micromagnetic}), yielding a saturation
          magnetization of
          $M_\mathrm{sat} = \SI{1.1 \pm
            0.1e6}{\ampere\per\meter}$. The inset shows a close-up
          plot of mode 1's frequency shift in the switching region
          around zero field for down- and up-sweep of the field (blue
          arrows). The jumps are highlighted by solid and dashed blue
          lines.\label{fig:DCM}}
\end{figure}

We probe the magnetic properties of each NW by measuring its
mechanical response to a uniform magnetic field $B$ up to
\SI{8}{\tesla} applied along its long axis.  In
particular, we measure the shift in the resonance frequency of each
flexural mode, $\Delta f_i = f_i - f_{0_i}$, as a function of $B$,
where $f_{0_i}$ is the resonance frequency at $B=0$.  Figure
\ref{fig:DCM} shows a typical measurement of the hysteretic response
of $\Delta f_1(B)$ and $\Delta f_2(B)$ carried out on NW 4 with
$T_{\text{bath}} = \SI{4.2}{\kelvin}$.  As in measurements of the
other NWs, the data show a smooth V-shaped response for most of the
field range, except for discontinuous inversions of the slope
(``jumps'') in reverse fields of around $\SI{\pm 40}{\milli\tesla}$.
These sharp features, which arise from the switching of the NW
magnetization, and the steady stiffening of the mechanical response as
$|B|$ increases are characteristic of a strong magnet with a square
magnetization hysteresis, whose easy axis is nearly parallel to the
applied field~\cite{gross_dynamic_2016}.  Therefore, the data point to
NWs with negligible magnetocrystalline anisotropy and an easy axis
coincident with their long axis, as set by the magnetic shape
anisotropy resulting from their extreme aspect ratio.

In order to extract specific magnetic properties from our
measurements, we compare them to micromagnetic simulations, which
model both the NW's magnetic state and the way in which its
interaction with $B$ affects the mechanical rigidity of the flexural
modes.  We use \textit{Mumax3}~\cite{vansteenkiste_design_2014,
	Exl2014}, which employs the Landau-Lifshitz-Gilbert micromagnetic
formalism with finite-difference discretization, together with
geometrical and material parameters to model each NW.  For a given
value of $B$ in a hysteresis loop, the numerical simulation yields the
equilibrium magnetization configuration and the total magnetic energy
$E_m$ corresponding to that configuration.  Just as in dynamic
cantilever magnetometry
(DCM)~\cite{gross_dynamic_2016,mehlin_observation_2018}, the frequency
shift of each flexural mode is proportional to the curvature of the
system's magnetic energy $E_m$ with respect to rotations $\theta_i$
corresponding to each mode's oscillation:
\begin{equation}
\Delta f_i = \frac{f_{0_i}}{2 k_i l_e^2} \left ( \left.  \frac{\partial^2 
	E_m}{\partial \theta_i^2} \right|_{\theta_i=0} \right ),
\label{eq:DCM}
\end{equation}
where $l_e$ is an effective length, which takes into account the shape
of the flexural mode~\cite{stipe_magnetic_2001}.  Therefore, by
numerically calculating the second derivatives of $E_m$ with respect
to $\theta_i$ at each $B$, we simulate $\Delta f_i (B)$ (see Appendix
\ref{app:micromagnetic}).  Note that, unlike in standard DCM, where
the magnetic sample is attached to the end of the cantilever, each NW
is magnetic along its full length.  Because of the mode shape,
different parts of the NW rotate by different angles during a flexural
oscillation.  This effect must be carefully considered in order to
correctly model the system.

\begin{figure}
	\centering
	\includegraphics[width=0.95\textwidth]{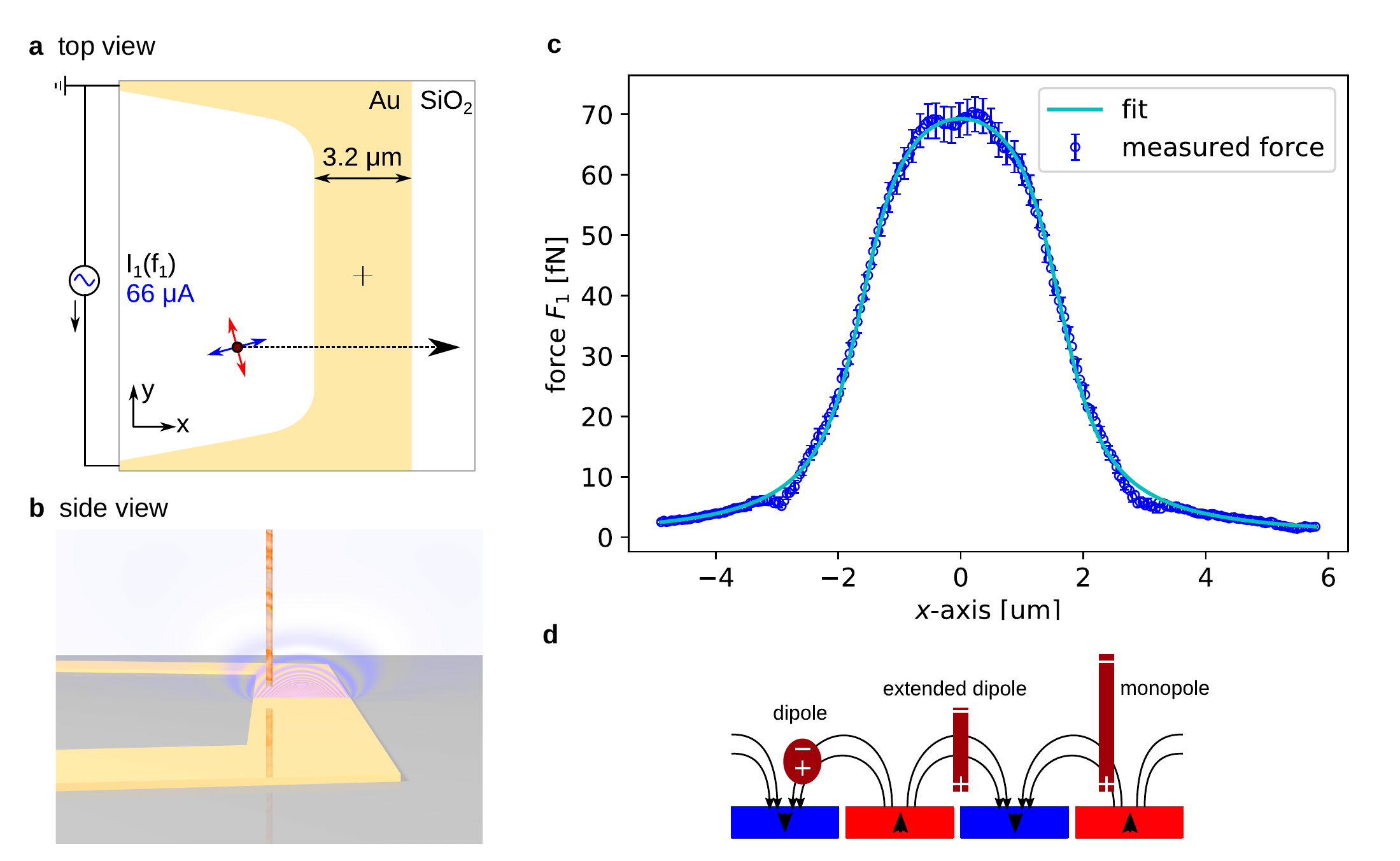}
	\caption{\textsf{\textbf{a}}) Sketch of
		the Au wire dimensions, NW mode directions, scanning
		direction and applied drive tone. \textsf{\textbf{b}}) Illustration of the scanning probe measurement. \textsf{\textbf{c}}) Measured response of the first mode of NW 4 as a function of position along the line scan indicated by the arrow shown in \textsf{\textbf{a}}) to a resonant drive field for a tip-sample distance of $d_z = \SI{200}{\nano\meter}$. The best fit according to equation \eqref{eq:monopole} is represented by the solid cyan line and yields a value of $q_0 = \SI{9.7 \pm 0.4e-9}{\ampere\meter}$. \textsf{\textbf{d}}) Illustration of three tip models for magnetic tips. The FEBID NWs correspond to the monopole tip model. Figure adapted from \cite{hug_quantitative_1998}. \label{fig:wireresponse}}
\end{figure}

The excellent agreement between the measured and simulated
$\Delta f_i (B)$ in Figure~\ref{fig:DCM} is typical for all measured
NWs.  For each NW, the mechanical parameters used in the simulation
are extracted from measurements of the thermal motion at $B = 0$,
while geometrical parameters are estimated from SEM images.  We adjust
the value of the saturation magnetization $M_\mathrm{sat}$ in order to
bring the curves into agreement, giving us a sensitive measurement of
this material property.  $M_\mathrm{sat}$ is found to be
$\SI{1.1 \pm 0.1e6}{\ampere\per\meter}$, where the uncertainty is
dominated by the estimation of the NW geometry from SEM.  This value
compares well with the expected Co composition of the FEBID NWs and
literature values for the saturation magnetization of Co, as well as
with electron holography measurements of similar NWs
\cite{pablo-navarro_tuning_2017}.  In addition, the simulations show
that the magnetization of the NWs is axially aligned in remanence for
up to about \SI{40}{\milli\tesla} of reverse field.  These results are
consistent with nanoSQUID measurements of similar Co NWs carried out
at 15 K by Mart\'inez-P\'erez et
al~\cite{martinez-perez_nanosquid_2018-1}.  This axially aligned
remanent magnetization can be represented by a magnetic charge
distribution in the form of an elongated dipole, leaving a
monopole-like distribution localized at the free end of the NW to
interact with an underlying sample, as shown in
Figure~\ref{fig:wireresponse}d.

\section{Measurement of magnetic field profile}
\label{sec:magfield}

In order to determine the behavior of the FEBID NWs as scanning
probes, we approach and scan a nearly planar sample with respect to NW
4 at $T_{\text{bath}}=\SI{4.2}{\kelvin}$.  The sample consists
of a \SI{6}{\micro\meter}-long, \SI{3.2}{\micro\meter}-wide, and
\SI{240}{\nano\meter}-thick Au wire patterned between two contact pads
on a Si substrate (Figure \ref{fig:wireresponse}a and b).  By passing a current through the wire, we produce
a well-known magnetic field profile $\mathbf{B}_{\text{AC}}(x,y,z)$
given by the Biot-Savart relation, with which we drive NW oscillations
and calibrate its response, as done in standard
MFM~\cite{lohau_quantitative_1999, schendel_method_2000,
	kebe_calibration_2004}.  By applying an excitation current
containing two sine waves, each at the frequency of one of the NW
modes $f_1$ and $f_2$, we drive the NW as we scan it across the Au
wire at a fixed tip-sample spacing.  Both the resonance frequencies
$f_i$ and oscillation amplitudes $r_i$ are tracked using two
phase-locked loops. The corresponding values of the force driving each
mode on resonance are calculated using $F_i = r_i k_i / Q_i$ (see Supplementary
Information).  Figure~\ref{fig:wireresponse}c shows the response of
mode 1 for a drive current amplitude of \SI{47}{\micro\ampere} as the
NW is scanned above the Au wire at a fixed distance
$d_z = \SI{200}{\nano\meter}$ in the absense of static magnetic field
($B = 0$). Since the first mode is nearly aligned with the
$x$-direction ($\alpha \approx \SI{7.3}{^\circ}$) and thus along the
direction of $\mathbf{B}_{\text{AC}}$, the orthogonal second mode
has almost no response to the driving tone at $f_2$ and is not shown.

From our torque magnetometry measurements, we know that the magnetic
NWs have an axially aligned remanent magnetization.  Because the decay
length of the magnetic field from our sample is much shorter than the
NW length, the sample fields only interact with the monopole-like
magnetic charge distribution at the free end of the
NW~\cite{hug_quantitative_1998}.  This charge distribution then
determines the NW's response to magnetic field profiles produced by a
sample.  For a monopole-like NW tip, we can relate the driving
magnetic field and the force it produces on the NW by
\begin{equation}
F_i = q_0 \mathbf{B}_{\text{AC}} \cdot \hat{\mathbf{r}}_i,
\label{eq:monopole}
\end{equation}
where $q_0$ is an effective magnetic monopole moment describing the
tip magnetization and $\hat{r}_i$ is the unit vector in the direction
of displacement of mode $i$.  In this point-probe approximation, we
consider the interaction of dipole and higher multipoles of the
magnetic charge with the driving field to be negligible.  As shown by
the agreement between the field calculated from the Biot-Savart law
and the measured response of NW 4 in Figure~\ref{fig:wireresponse}c,
this approximation is valid for our NWs.  Control experiments, using
the applied magnetic field to initialize the NW magnetization along
the opposite direction also show that spurious electrostatic driving
of the NW modes is negligible. Combining measurements at different
$d_z$ and different driving currents, we find that NW 4 has an
effective magnetic charge of
$q_0 = \SI{9.7 \pm 0.4e-9}{\ampere\meter}$.  Given our thermally
limited force sensitivity of
$\SI{25}{\atto\newton}/\sqrt{\SI{}{\hertz}}$ at
$T_{\text{bath}} = \SI{4.2}{\kelvin}$, this value of $q_0$ gives our
sensors a sensitivity to magnetic field of around
$\SI{3}{\nano\tesla}/\sqrt{\SI{}{\hertz}}$.  This sensitivity is
similar to those of some of the most sensitive scanning probes
available, including scanning NV magnetometers and scanning SQUIDs.

Furthermore, the magnetic charge model allows us to estimate the stray
field and field gradients produced by the NW tip, so that we can
assess its potential for perturbing the magnetic state of the sample
below.  At a distance of $d_z = \SI{50}{\nano\meter}$ from the NW tip,
the stray magnetic field and magnetic field gradients are
$B_\mathrm{tip} = \mu_0 q_0 /(4\pi z_\mathrm{mono}^2) \approx
\SI{60}{\milli\tesla}$ and
$dB_\mathrm{tip}/dz \approx \SI{1}{\mega\tesla\per\meter}$.  The stray
field is of similar size to that produced by a conventional MFM
tip~\cite{zhu_modern_2005}.  For future NW devices to be less
invasive, i.e.\ having less magnetic charge at their tips, sharper
tips than those produced here, which are more than 100 nm in diameter,
will be required~\cite{berganza_observation_2018}.  The large magnetic
field gradients, however, combined with the NWs' excellent force
sensitivity may make the NWs well-suited as transducers in sensitive
magnetic resonance force microscopy~\cite{poggio_force-detected_2010}.

\begin{figure}
	\centering
	\includegraphics[width=0.95\textwidth]{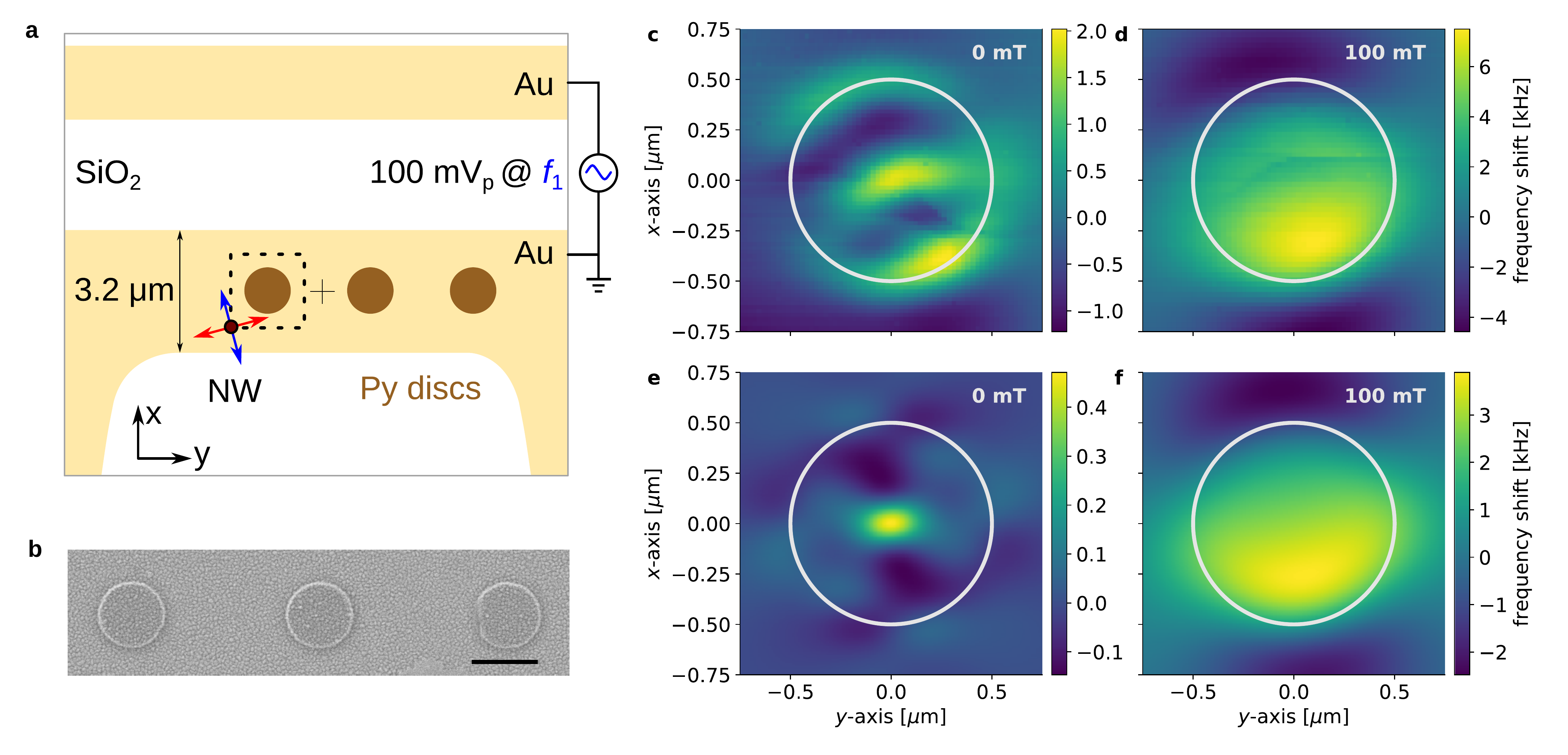}
	\caption{\textsf{\textbf{a}}) Sketch of the Au wire sample
		with three \SI{1}{\micro\meter} wide permalloy disks
		patterned on top. The NW is driven resonantly by applying a
		gate voltage of \SI{100}{\milli\volt_p} amplitude at
		$f_1$. \textsf{\textbf{b}}) SEM image of the permalloy
		disks. The scale-bar is \SI{1}{\micro\meter}.
		\textsf{\textbf{c}}), \textsf{\textbf{d}}) NW-MFM images using NW 4 with an external magnetic field applied perpendicular to the disk plane of $B=0$ and $B=\SI{100}{\milli\tesla}$, respectively. \textsf{\textbf{e}}), \textsf{\textbf{f}}) corresponding simulations of the expected contrast obtained by applying the monopole model \eqref{eq:monopole} to the disk's stray field calculated by \textit{Mumax3}. The disk circumference is highlighted by a white circle. \label{fig:2Dimage}}
\end{figure}

\section{Magnetic field imaging}
\label{sec:imaging}

In Figure~\ref{fig:2Dimage}, we demonstrate the ability to image
sub-micrometer features using a FEBID NW as MFM sensor on a permalloy
disk (Ni$_{0.81}$Fe$_{0.19}$) at
$T_{\text{bath}} = \SI{4.2}{\kelvin}$. Three disks,
\SI{1}{\micro\meter} in diameter and \SI{40}{\nano\meter}-thick, are
patterned on top of the Au wire, of which one is imaged. During the
scan, the first flexural mode of the NW is electrically driven on
resonance using an AC voltage applied between the Au wire and a third
lead (Figure \ref{fig:2Dimage}a). Frequency shift and dissipation are
recorded using a phase-locked loop.  Figures~\ref{fig:2Dimage}c and d
show a $1.5 \times 1.5$~\SI{}{\micro\meter^2} image of frequency shift
$\Delta f_1(x,y)$, as the disk is scanned below NW 4 for
$d_z = \SI{100}{\nano\meter}$ at $B = 0$ and for
$d_z = \SI{150}{\nano\meter}$ at $B = \SI{100}{\milli\tesla}$,
respectively. At $B = 0$ (Figure \ref{fig:2Dimage}c), the
magnetization of the disk is arranged in a remanent vortex
configuration~\cite{badea_magneto-optical_2015}, as verified by a
micromagnetic simulation carried out with \textit{Mumax3}. Figure
\ref{fig:2Dimage}e shows the magnetic image contrast expected from the
simulation using the monopole model \eqref{eq:monopole}, in which the
frequency shift of the NW mode is proportional to the stray field
derivative along the mode direction. While the contrast measured at
the edges of the disk is due to topographic features, the contrast in
the center is consistent with what is expected from the stray field of
a vortex core. The image taken at $B=\SI{100}{\milli\tesla}$, Figure
\ref{fig:2Dimage}d, shows an almost homogeneous magnetic imaging
contrast across the disk. The corresponding simulation in Figure
\ref{fig:2Dimage}f agrees well with the measurement and reveals that
while the vortex core is still present in the center of the disk, its
stray field is overshadowed by the field originating from the outer
parts of the disk, where the magnetization tilts out of plane. The
region of high frequency shift in the bottom right quadrant of Figure
\ref{fig:2Dimage}d is explained in the simulation by assuming a small
tilt ($\approx \SI{1}{^\circ}$) of the disk plane with respect to the
external field, resulting in the vortex core being slighltly offset
from the center of the disk.  Although a detailed interpretation of
the NW MFM images and a quantitative comparison of the measured and
calculated frequency shifts is beyond the scope of this work, they
showcase the high sensitivity and potential spatial resolution of the
FEBID NWs transducers.

\section{Conclusion}
\label{sec:conc}

In the past, FEBID-grown NWs have been patterned directly on tips of
atomic force microscopy (AFM) cantilevers in an effort to improve
spatial resolution~\cite{utke_high-resolution_2002,
	stiller_functionalized_2017, berganza_observation_2018}.  Our
results make clear that such nanocrystalline metallic NWs can have
surprisingly high mechanical quality, making FEBID a promising and
versatile method for producing nanometer-scale force transducers.  In
principle, a FEBID NW patterned on the tip of a standard AFM or MFM
cantilever could be used to add sensitive 2D lateral force and
dissipation detection capabilities.  Such a system would be capable of
vectorial force sensing in 3D, i.e.\ mapping both the size and the
direction in 3D of tip-sample forces.

In addition to demonstrating the high-force sensitivity of FEBID-grown
NWs, we also show their excellent magnetic properties.  The Co NWs
measured here maintain a saturation magnetization, which is 80\% of
the value of pure Co.  They also have an axially aligned remanent
magnetization with a switching field around \SI{40}{\milli\tesla}.
These magnetic properties, combined with the aforementioned mechanical
properties, make these NWs among the most sensitive sensors of local
magnetic field.  The ability to fine tune the NW geometry, especially
making them thinner and sharper, may allow for even better field
sensitivities and spatial resolutions in the future.  NW MFM with such
transducers may prove ideal for investigating subtle magnetization
textures and current distributions on the nanometer-scale, which -- so
far -- have been inaccessible by other methods.

\appendix
\section{Interferometric Measurement of Flexural Motion}
\label{app:interferometer}

We use a custom-built interferometer to detect the thermal
or driven motion of the NW of interest. At its heart is a four arm
fiber coupler with a 95:5 coupling ratio. A Toptica
\SI{1550}{\nano\meter} wavelength laser is connected to the input port
and can be attenuated to the desired power. At the experiment arm of
the fiber coupler ($\approx 5\%$ transmission), an objective with a
high numerical aperture focuses the laser light with a beam waist of
around \SI{1.65}{\micro\meter} onto the NW. The interferometric
displacement signal arises from the weak cavity between the NW and the
end facet of the optical fiber. At the signal port of the fiber
coupler the signal is converted to a voltage by a Femto OE-300
photoreceiver and subsequently split into its DC and AC parts. Knowing
the wavelength of the laser, the conversion factor between
displacement and voltage can be determined accurately and is usually
in the range of a few \si{\micro\meter\per\volt}. A more detailed
description of the optical setup can be found in the supplementary
information of Ref. \cite{rossi_magnetic_2019}.

The thermal displacement noise PSD, which is the projection of the
motion of the two first order flexural modes onto the direction of the
local optical gradient, can be described by the
fluctuation-dissipation theorem following the derivations in
Ref. \cite{braakman_force_2019} as,
\begin{equation}
S_{xx}(\omega) = u^2(z) \frac{4 k_B T}{m_\mathrm{eff}} \left( \frac{\omega_1}{Q_1} \frac{\cos^2 \alpha}{\left(\omega^2 - \omega_1^2\right)^2 + \left( \frac{\omega \omega_1}{Q_1}\right)^2} + \frac{\omega_2}{Q_2} \frac{\sin^2 \alpha}{\left(\omega^2 - \omega_2^2\right)^2 + \left( \frac{\omega \omega_2}{Q_2}\right)^2} \right) + S_\mathrm{n}
\label{eq:psd}
\end{equation}
where $u(z)$ is the mode shape, $k_B$ the Boltzmann constant,
$T_\mathrm{eff}$ and $m_\mathrm{eff}$ the effective temperature and
mass of the NW resonator, $\omega_{1, 2}$ and $Q_{1, 2}$ the resonance
frequencies and quality factors of the two first order flexural modes,
$\alpha$ the measurement angle between the direction of the first mode
and the optical gradient, and $S_\mathrm{n}$ the background
noise. Depending on the position of the NW inside the beam waist, the
optical gradient direction can be chosen at will. Ideally, however, it
is aligned with the optical axis in order to achieve the best
signal-to-noise ratio. The $z$-direction is aligned with the NW axis
and its origin is located at the base of the NW. From the fit
parameters, the spring constants
$k_{1, 2} = m_\mathrm{eff} \omega_{1, 2}^2$ and the thermally limited
force sensitivity
$F_\mathrm{min} = \sqrt{4 k_B T m_\mathrm{eff} \omega / Q}$ can be
calculated.

\section{Micromagnetic Simulations}
\label{app:micromagnetic}

The principles of simulating the torque magnetometry signal with micromagnetic solvers are described in
Refs. \cite{gross_dynamic_2016, mehlin_stabilized_2015,
	rossi_magnetic_2019}. In these works, it is only the tip of the
mechanical resonator which is magnetic, therefore the system can be
modelled as a magnetic object oscillating in a homogeneous external
magnetic field. The mode shape of the mechanical resonator enters into
the calculation only in the form of the effective length, simplifying
the mechanics to that of a harmonic oscillator.  For the Co NWs, which
are both the mechanical resonator and the magnetic object, the mode
shape has to be taken into account. Each longitudinal segment of the
NW rotates by a different angle during a flexural oscillation,
experiencing a different tilt of the external magnetic field. We
account for this effect by applying a spatially dependent external
field in the simulation, rather than altering the geometry, which is
impractical. For positive (negative) deflections in experiment, the
tilt direction of the field in the simulations increases (decreases)
with the $z$ position along the NW. The magnitude of the tilt angle
follows the Euler-Bernoulli equation, reflecting the mode shape. We
choose a maximum oscillation amplitude (at the tip) large enough to
account for the finite precision of the simulation. The torque signal
can then be calculated using the magnetic energy of the system for
small positive, negative, and no deflection given by the simulation
combined with a finite difference approximation for the second
derivative in equation \eqref{eq:DCM}~\cite{mehlin_observation_2018}.

The final geometry used in the simulation of Figure \ref{fig:DCM} is a
$\SI{9.1}{\micro\meter}$ long, elliptic cylinder, whose diameters are
modulated along the $z$ direction, as determined from the SEM
images. The average diameters along the two mode directions are
$d_1 = \SI{135}{\nano\meter}$ and $d_2 = \SI{125}{\nano\meter}$. Space
is discretized to $\SI{5}{\nano\meter}$. Material parameter values are
the saturation magnetization
$M_{sat} = \SI{1.1 \pm 0.1e6}{\ampere\per\meter}$ and the exchange
stiffness $A_{ex} = \SI{38}{\pico\joule\per\meter}$.  The latter has
been chosen to match the switching field of the NW, and is
significantly larger than other values reported for Co
\cite{shirane_spin_1968, vernon_brillouin_1984, krishnan_fmr_1985,
	liu_exchange_1996}. This discrepancy arises because the switching
field also depends sensitively on geometrical and material
imperfections, which we do not attempt to model.  Nevertheless,
control simulations confirm that the overall magnetization reversal
process and the remanent states are unaffected by such differences in
$A_{ex}$.

In an effort to determine the effect of nanocrystallinity in the NW,
we have run simulations with NW divided into grains of around
$\SI{10}{\nano\meter}$ size, giving each grain a uniaxial anisotropy
with $K_1 = \SI{530}{\kilo\joule\per\cubic\meter}$ and a random
orientation of the anisotropy axis. We find that this refinement does
not significantly change the simulation results with respect to
standard simulations assuming homogeneous material without crystalline
anisotropy.

\begin{acknowledgments}
We thank Sascha Martin and his team in the machine shop of the
Physics Department at the University of Basel for help building the
measurement system.  We acknowledge the support of the Kanton
Aargau, the ERC through Starting Grant NWScan (Grant 334767), the
SNF under Grant 200020-178863, the Swiss Nanoscience Institute, and
the NCCR Quantum Science and Technology (QSIT) as well as from the Spanish Ministry of Economy and Competitiveness (MINECO) through the projects MAT2017-82970-C1 and MAT2017-82970-C2 and from the Aragon Regional Government (Construyendo Europa desde Aragón) through project E13\_17R, with European Social Fund funding. J. P.-N. grant is funded by the Ayuda para Contratos Predoctorales para la Formación de Doctores, Convocatoria Res. 05/06/15 (BOE 12/06/15) of the Secretaría de Estado de Investigación, Desarrollo e Innovación in the Subprograma Estatal de Formación of the Spanish Ministry of Economy and Competitiveness with the participation of the European Social Fund.
\end{acknowledgments}

%

\end{document}



\title{Supplementary Information: Nanowire magnetic force sensors fabricated by focused electron beam induced deposition}



\author{H.~Mattiat} \affiliation{Department of Physics, University of
	Basel, 4056 Basel, Switzerland} \affiliation{Swiss Nanoscience
	Institute, University of Basel, 4056 Basel, Switzerland}

\author{N.~Rossi} \affiliation{Department of Physics, University of
	Basel, 4056 Basel, Switzerland} \affiliation{Swiss Nanoscience
	Institute, University of Basel, 4056 Basel, Switzerland}

\author{B.~Gross} \affiliation{Department of Physics, University of
	Basel, 4056 Basel, Switzerland} \affiliation{Swiss Nanoscience
	Institute, University of Basel, 4056 Basel, Switzerland}

\author{J.~Pablo-Navarro} \affiliation{Laboratorio de Microscop\'ias
	Avanzadas (LMA), Instituto de Nanociencia de Arag\'on (INA),
	Universidad de Zaragoza, 50018 Zaragoza, Spain}
\affiliation{Instituto de Ciencia de Materiales de Arag\'on and
	Departamento de F\'isica de la Materia Condensada, CSIC-Universidad
	de Zaragoza, 50009 Zaragoza, Spain}

\author{C.~Mag\'{e}n} \affiliation{Laboratorio de Microscop\'ias
	Avanzadas (LMA), Instituto de Nanociencia de Arag\'on (INA),
	Universidad de Zaragoza, 50018 Zaragoza, Spain}
\affiliation{Instituto de Ciencia de Materiales de Arag\'on and
	Departamento de F\'isica de la Materia Condensada, CSIC-Universidad
	de Zaragoza, 50009 Zaragoza, Spain}

\author{R.~Badea} \affiliation{Department of Physics, Case Western
	Reserve University, Cleveland, Ohio 44106, USA}

\author{J.~Berezovsky} \affiliation{Department of Physics, Case
	Western Reserve University, Cleveland, Ohio 44106, USA}

\author{J.~M.~De Teresa} \affiliation{Laboratorio de Microscop\'ias
	Avanzadas (LMA), Instituto de Nanociencia de Arag\'on (INA),
	Universidad de Zaragoza, 50018 Zaragoza, Spain}
\affiliation{Instituto de Ciencia de Materiales de Arag\'on and
	Departamento de F\'isica de la Materia Condensada, CSIC-Universidad
	de Zaragoza, 50009 Zaragoza, Spain}

\author{M.~Poggio} \affiliation{Department of Physics, University of
	Basel, 4056 Basel, Switzerland} \affiliation{Swiss Nanoscience
	Institute, University of Basel, 4056 Basel, Switzerland} \email{martino.poggio@unibas.ch}


\date{\today}
\maketitle

\section{Bolometric heating}
\label{app:heating}

In order to investigate any bolometric heating of the nanowires (NWs)
due to the readout laser, the thermal noise power spectral density
(PSD) is measured at the tip of NW 4 for laser powers from 0.5 to
\SI{25}{\micro\watt} in room temperature
($T_{\text{bath}}=\SI{293}{\kelvin}$), liquid nitrogen
($T_{\text{bath}}=\SI{77}{\kelvin}$), and liquid helium environments
($T_{\text{bath}}=\SI{4.2}{\kelvin}$). In addition, the influence of
the laser beam's position along the NW's long axis is examined for a
fixed laser power of \SI{20}{\micro\watt}. All laser powers provided
in this section correspond to the total power at the output arm of the
interferometer at the bottom of the probe, not the fraction of the
power incident on the NW.

Generally, for high laser powers, a negative resonance frequency shift
and a high thermal noise amplitude are observed, corresponding either
to an increase in NW temperature or to a decrease in effective mass of
the resonator. Since the latter can be excluded, we fix the effective
mass to \SI{2.6 \pm 0.2 e-16}{\kilogram}, as extracted from the
highest observed values at small incident laser powers with
$T_{\text{bath}}=\SI{293}{\kelvin}$. The remaining data is then fit to
equation (3) in the main text with the NW temperature $T = T_\mathrm{NW}$ as a free
parameter.

The changes observed in the mechanical properties of the NW as a
function of laser power and focal position are a consequence of the
temperature dependence of the NW's Young's modulus
\cite{gil-santos_optical_2013}. A change of the Young's modulus is
reflected in a change of a resonators resonance frequency,
specifically in a negative frequency shift for higher NW temperatures.

\subsection{Laser power dependence}

\begin{figure}
	\centering
	\includegraphics[width=0.9\textwidth]{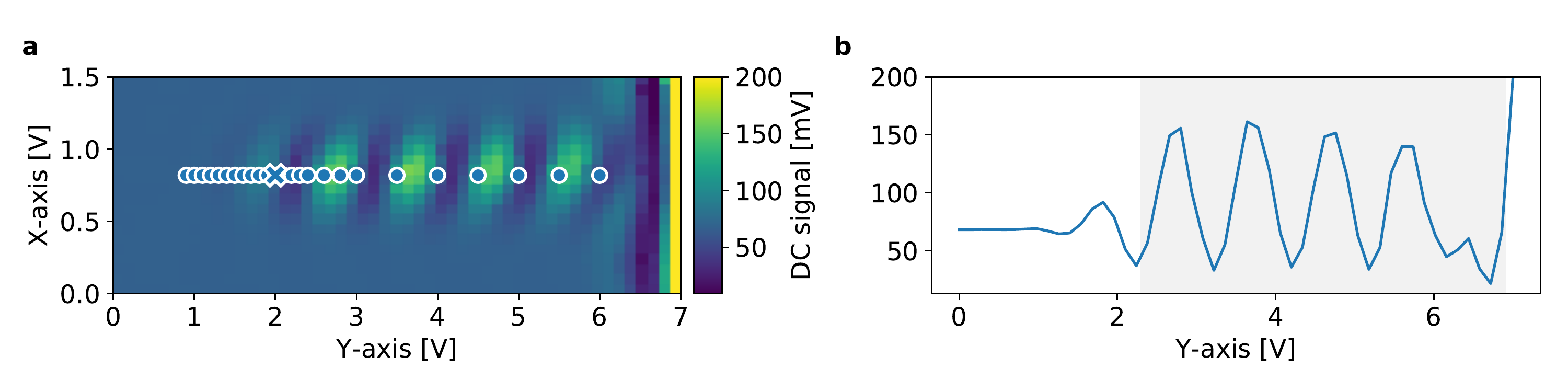}
	\caption{\textsf{\textbf{a}}) Scanning interferometer image of
		NW 4 at $T_{\text{bath}}=\SI{4.2}{\kelvin}$, with the spots
		marked at which the laser power dependence measurements are
		taken. The position marked by a cross (X) corresponds to the
		measurement position for the data shown in Figures
		\ref{fig:app_powerdep} and
		\ref{fig:app_powerdep_freq_vs_T}. Points correspond to
		measurement positions for data shown in Figure
		\ref{fig:app_freq_Teff_axis}. \textsf{\textbf{b}}) Line scan
		along the long axis of the NW. The gray box highlights the
		extent of the NW.  Interference fringes are visible in the
		images because the $z$-piezo scanner movement has a small
		spurious component along the $x$-direction.}
	\label{fig:app_meas_points}
\end{figure}

\begin{figure}
	\centering
	\includegraphics[width=1.0\textwidth]{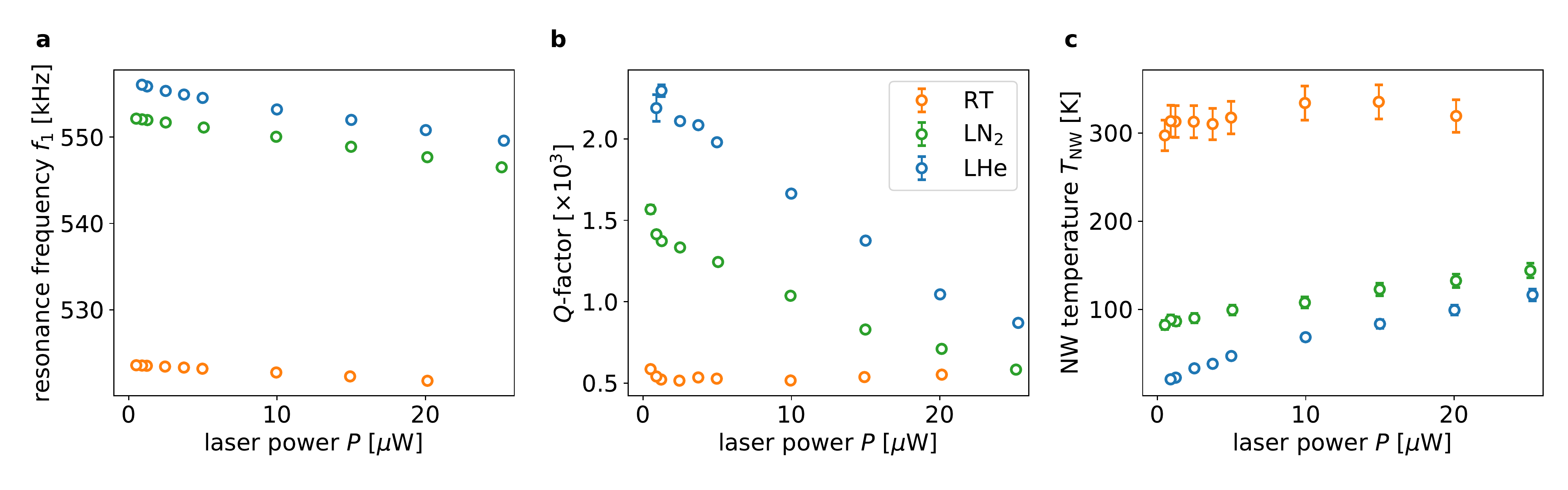}
	\caption{Resonance frequency \textsf{\textbf{a}}), quality
		factor \textsf{\textbf{b}}) and NW temperature
		\textsf{\textbf{c}}) at $T_{\text{bath}}=\SI{293}{\kelvin}$
		(orange), \SI{77}{\kelvin} (green), and \SI{4.2}{\kelvin}
		(blue) for different laser powers.  These power dependences
		are measured with the laser aligned to the tip of the NW,
		corresponding to the cross (X) in
		Figure~\ref{fig:app_meas_points} a. }
	\label{fig:app_powerdep}
\end{figure}

\begin{figure}
	\centering
	\includegraphics[width=0.8\textwidth]{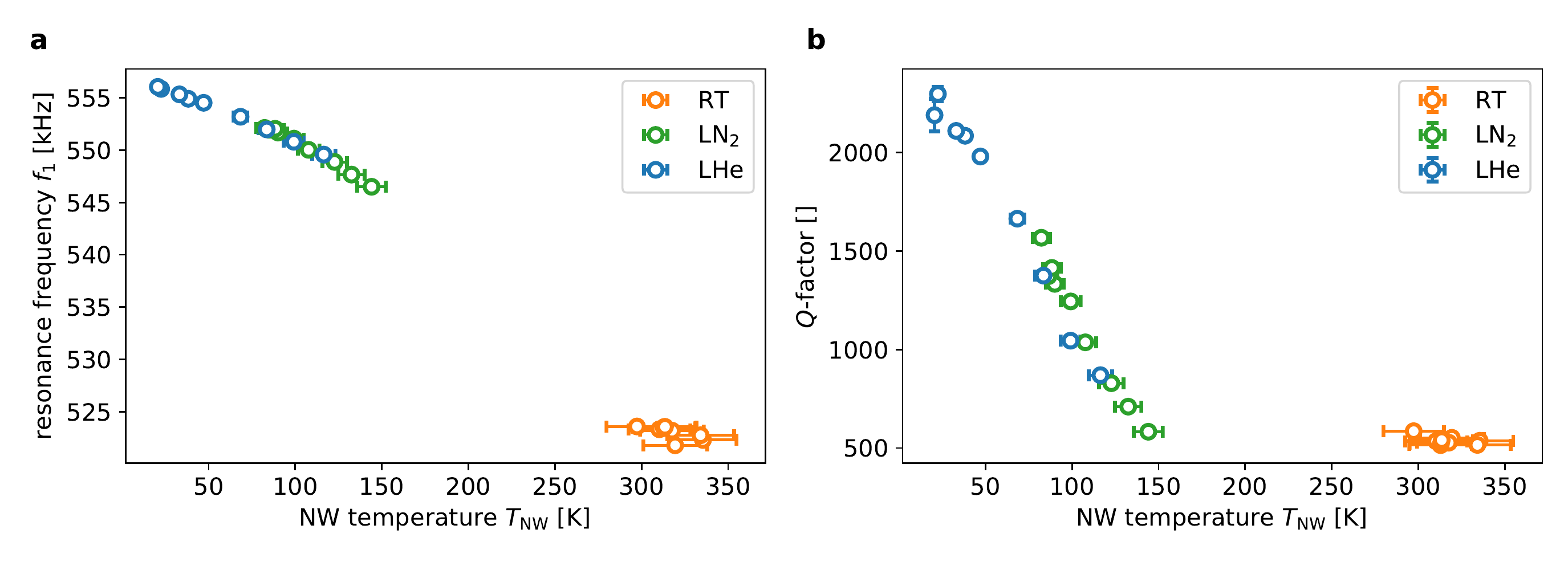}
	\caption{\textsf{\textbf{a}}) The resonance frequency of NW 4
		plotted against the NW temperature for the full data set at
		all three different bath temperatures. The overlap of the
		points, especially at $T_{\text{bath}}=\SI{77}{\kelvin}$ and
		\SI{4.2}{\kelvin}, points to a bolometric heating
		effect. \textsf{\textbf{b}}) Quality factor $Q$ against the
		NW temperature.  These measurments are carried out with the
		laser aligned to the tip of the NW, corresponding to the
		cross (X) in Figure~\ref{fig:app_meas_points} a.}
	\label{fig:app_powerdep_freq_vs_T}
\end{figure}

Figure \ref{fig:app_meas_points}a shows an image of NW 4 with the
positions marked, at which the thermal noise PSD is measured as a
function of laser power while the bath is held at
$T_{\text{bath}}=\SI{4.2}{\kelvin}$. Figure \ref{fig:app_meas_points}b
shows a line cut of this image along the NW's long axis with the gray
area highlighting the spatial extent of the NW.  The NW's length is
used to calibrate the conversion between piezo scanner voltage and its
displament in \SI{}{\micro\meter}.  Resonance frequencies, quality
factors, and NW temperatures of the first mode are measured as a
function of laser power at the position marked with the cross (X) in
Figure \ref{fig:app_meas_points} and plotted in Figure
\ref{fig:app_powerdep}. For all three bath temperatures, the resonance
frequency shows a linear decrease with increasing laser power. The
quality factor exhibits a similar drop as a function of power for both
$T_{\text{bath}}=\SI{4.2}{\kelvin}$ and
$T_{\text{bath}} = \SI{77}{\kelvin}$, whereas at
$T_{\text{bath}}=\SI{293}{\kelvin}$ it appears constant within the
error. The NW temperature increases with laser power at
$T_{\text{bath}}=\SI{4.2}{\kelvin}$ and
$T_{\text{bath}} = \SI{77}{\kelvin}$, while at
$T_{\text{bath}}=\SI{293}{\kelvin}$, it appears roughly constant.

\subsection{Focal position dependence}

\begin{figure}
	\centering
	\includegraphics[width=0.8\textwidth]{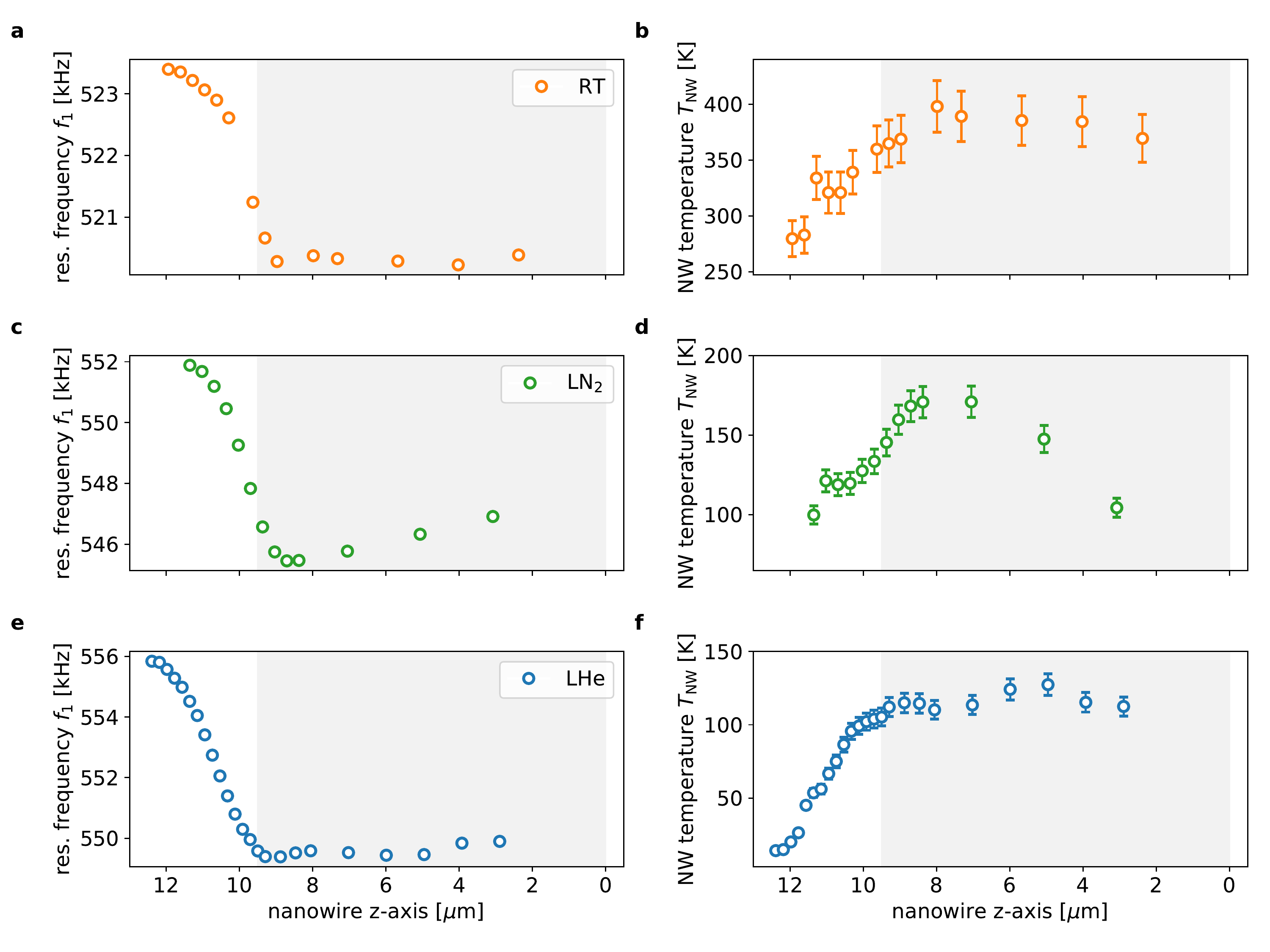}
	\caption{\textsf{\textbf{a}})--\textsf{\textbf{f}}) Resonance
		frequency and NW temperature at different points along the
		long axis of NW 4. The gray boxes mark the spatial extent of
		the NW. Moving the NW into the laser results in a lowering
		of the resonance frequency since the amount of incident
		power increases on the wire, thus heating it up. Once the NW
		is fully immersed in the laser beam, the NW temperature is
		roughly constant. The drop for the data at
		$T_{\text{bath}}=\SI{77}{\kelvin}$ in \textsf{\textbf{d}})
		can be explained by a small piezo drift perpendicular to the
		wire axis. Close to the base point the NW temperature
		decreases again due to a portion of the laser light being
		scattered at the chip edge. \textit{Note}: The abscissa is
		inverted and its origin corresponds to the base point of the
		NW.}
	\label{fig:app_freq_Teff_axis}
\end{figure}

For a second set of measurements, the focal spot is moved along the
$z$-axis (long axis) of the NW.  Thermal noise PSDs are measured at
the points shown in Figure \ref{fig:app_meas_points}.  Figure
\ref{fig:app_freq_Teff_axis} shows the resonance frequency and the NW
temperature along the NW, extracted from thermal noise PSDs, while
taking into account the mode shape $u(z)$. As the focal spot is moved
onto the NW tip and begins to overlap with the NW, the resonance
frequency shifts downwards and the NW temperature rises, since maximal
power impinges on the NW. At the other end of the NW, close to its
base, the laser light is cut off by scattering at the chip
edge. Consequently the resonance frequency shifts up by a small amount
and the NW temperature is lowered. Since the signal in this region is
very small due to the mode's small displacement in this region and
scattering at the chip edge, some measurement points for the NW
temperature are omitted.

In the middle region of the NW, where the laser focal spot fully
overlaps with the NW, we measure a roughly constant NW temperature
profile.  This behavior is an indication that radiative losses -- not
thermal contact to the substrate -- are the dominant channel through
which the NW reaches thermal equilibrium.  This observation indicates
that the NW is well-isolated thermally from the substrate and
therefore, through the use of exchange gas to create thermal contact
to a sample, could be used as a sensitive thermal scanning
probe~\cite{halbertal_nanoscale_2016}.

\section{Biot-Savart field of wire and monopole model}
\label{app:Biot-Savart}

\begin{figure}
	\includegraphics[width=0.8\textwidth]{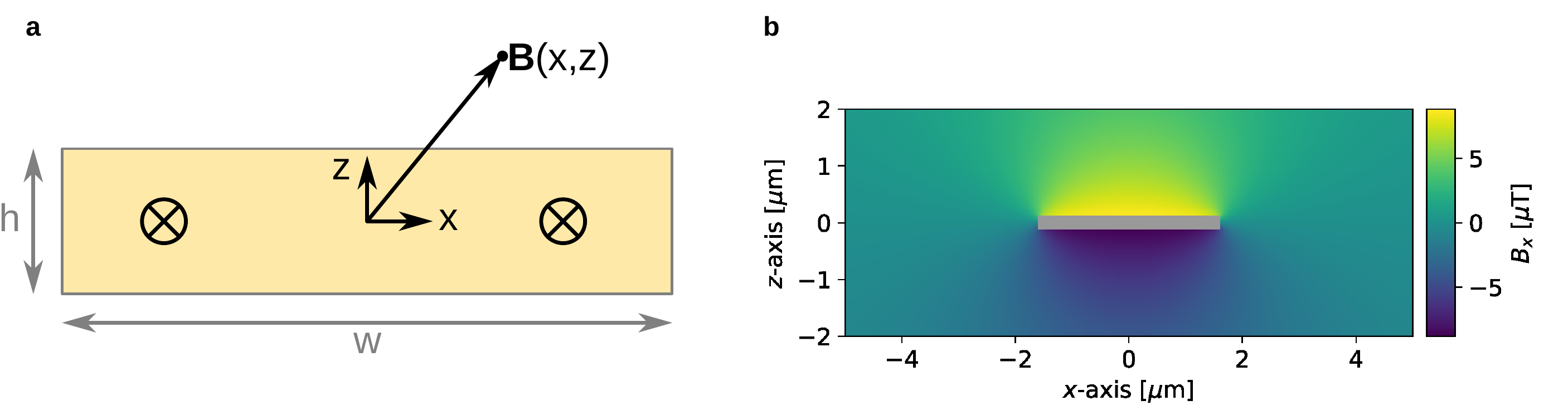}
	\caption{\textsf{\textbf{a}}) Cross section of an infinitely long
		rectangular conductor with width $w$ and height
		$h$. \textsf{\textbf{b}}) Plot of $B_x(x, z)$ according to
		equation \eqref{eq:app_Bx} for $w=\SI{3.2}{\micro\meter}$,
		$h=\SI{240}{\nano\meter}$ and a current of
		$I=\SI{47}{\micro\ampere}$.}
	\label{fig:rect_conductor}
\end{figure}
The magnetic field of an infinitely long conductor with rectangular
cross section of width $w$, height $h$ and a uniform current
$I=j_y w h$ flowing in $y$-direction (see Figure
\ref{fig:rect_conductor}), where $j_y$ is the current density along
the $y$-direction, can be calculated using the two-dimensional
Biot-Savart law
\begin{align}
B_x(x,z) &= \frac{\mu_0 I}{2 \pi wh} \int_{s} \frac{z'-z}{(x'-x)^2 + (z'-z)^2}dS'\\
B_z(x,z) &= \frac{\mu_0 I}{2 \pi wh} \int_{s} \frac{x'-x}{(x'-x)^2 + (z'-z)^2}dS'
\end{align}
where $S$ is its cross-sectional area. Solving the surface integrals
with the center of the coordinate system aligned with the geometrical
center of the conductor yields (see also Refs.\
\cite{goddenhenrich_probe_1990, kebe_calibration_2004})
\begin{align}
B_x(x,z) = \frac{\mu_0 I}{4 \pi w h} 
\left[(w/2-x) \log \frac{(w/2-x)^2 + (h/2-z)^2}{(w/2-x)^2 + (h/2+z)^2} \right. \nonumber\\
{} + (w/2+x) \log \frac{(w/2+x)^2 + (h/2-z)^2}{(w/2+x)^2 + (h/2+z)^2} \nonumber\\
{} + 2(h/2 - z)\left( \arctan \frac{w/2-x}{h/2-z} +  \arctan \frac{w/2+x}{h/2-z} \right) \nonumber \\
\left. {} - 2(h/2 + z)\left( \arctan \frac{w/2-x}{h/2+z} +  \arctan \frac{w/2+x}{h/2+z} \right)\right]
\label{eq:app_Bx}
\end{align}
and a similar expression for $B_z(x,z)$.

In order to extract a value for the monopole $q_0$, the response of
the NW to the driving tone is converted to a force using
$F_i = r_i k_i / Q_i$ ($i=1,2$ representing the two modes).  The
interferometric calibration factor between units of signal amplitude
in volts and displacement in meters and the measurement position along
the wire axis have to be taken into account. We then use equation
\eqref{eq:app_Bx} and the monopole model to fit the measured force
response of NW 4 by assuming
$\mathbf{B}_{\text{AC}}(x,y,z) = (B_x(x, z), 0, B_z(x, z))$ giving the
fit function for the first mode
\begin{equation}
F_1 = q_0 B_x(x, z) \cos \alpha'
\end{equation}
where $\alpha'$ is the angle between $x$-axis and the NWs first mode
direction. It differs from $\alpha$ by the angle between the
measurement direction and the optical axis, which is usually within
the range of \SI{2}{^\circ} when centered on the wire. The optical
axis is assumed to be aligned with the $x$-axis. The free parameters
of the fit are the value of the monopole $q_0$ and its effective
distance $z = z_\mathrm{mono}$. A typical fit is shown in Figure 4 of
the main text for a distance between the tip and Au wire of
$d_z = \SI{200}{\nano\meter}$ and a driving current of
\SI{47}{\micro\ampere}. The field profile fitting to the response
curve locates the effective monopole not at the tip of the NW but
higher along its axis at
$z_\mathrm{mono} = \SI{472 \pm 5}{\nano\meter}$.

\begin{figure}
	\centering
	\includegraphics[width=0.8\textwidth]{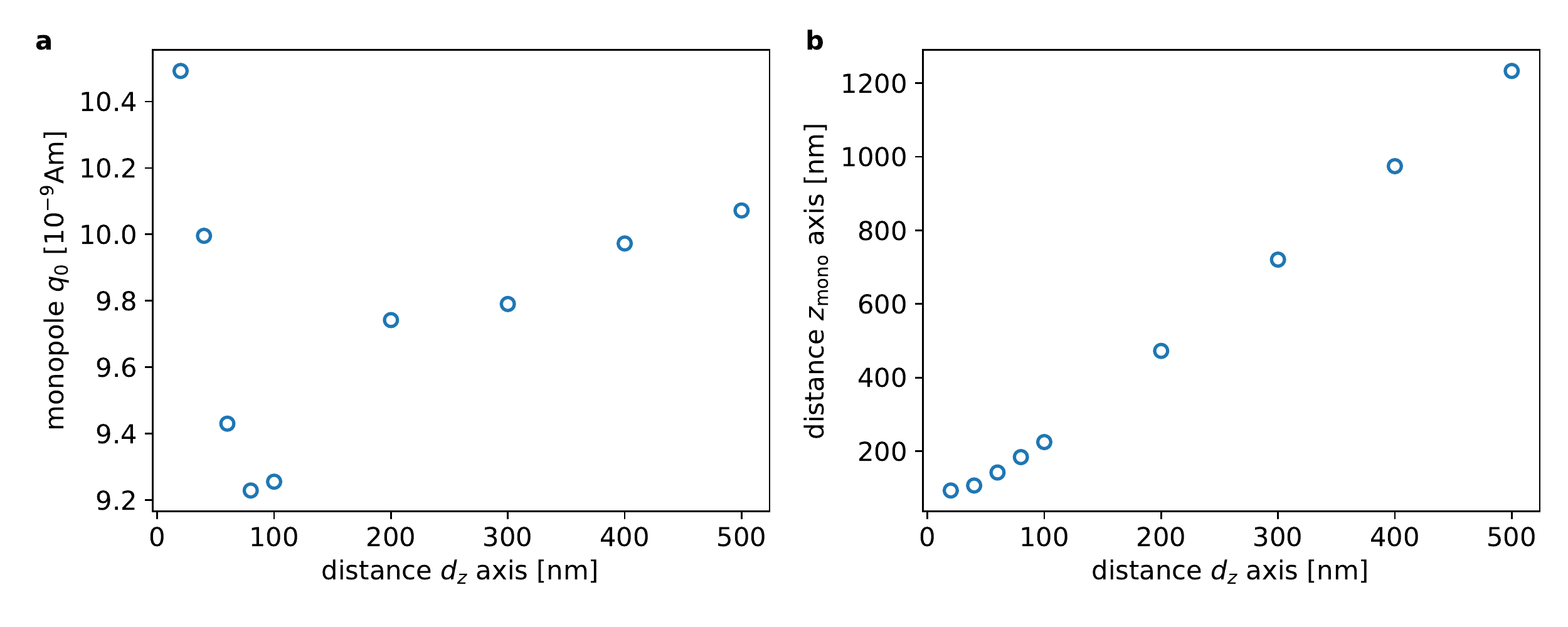}
	\caption{\textsf{\textbf{a}}) Value of the monopole $q_0$ and
		\textsf{\textbf{b}}) monopole distance $z_\mathrm{mono}$
		plotted against the tip-sample distance $d$. The monopole
		distance increases with the tip-sample distance. The fitted
		value of the monopole is scattered around a value of \SI{9.7
			\pm 0.4e-9}{\ampere\meter} and seems to change
		substantially below a tip-sample distance closer than the
		NWs diameter.}
	\label{fig:app_monopole_0T}
\end{figure}

This behavior is illustrated in more detail in Figure
\ref{fig:app_monopole_0T}. The value of the monople is scattered, for
different tip-sample distances, in a region around
$\SI{9.7 \pm 0.4e-9}{\ampere\meter}$. This spread can be related, on
the one hand, to imperfections of the Au-wire structure resulting in a
different field profile and, on the other hand, to the approximate
nature of the monopole model used for fitting the data. The tip of the
NW, which interacts with the Biot-Savart field of the Au wire, is an
extended object.  Depending on the tip-sample distance $d_z$, its
effective volume of interaction changes, leading to different values
for $q_0$ and $z_\mathrm{mono}$. In addition, a small tilt of the NW
from the $xy$-plane can also add error, since it introduces
sensitivity to the $z$-component of
$\mathbf{B}_{\text{AC}}(x,y,z)$. It should also be noted that, while
the zero-point of the tip-sample distance is precisely determined by a
soft touch of the NW tip onto the Au wire, the open-loop piezo
translation stage regulating the tip-sample distance in $z$-direction
is subject to piezoelectric creep.  We do not consider the
nonlinearity that creep introduces to the positioning of the NW.
Furthermore, the nature of the open loop scanners also introduces a
systematic error of about 5 \% in the conversion of the $x$- and
$y$-coordinates from voltage to \si{\micro\meter}, depending on the
calibration method used.

